\documentclass[a4paper, 12pt]{article}
\usepackage[dvips]{graphics}
\begin{document}

\title{\huge  Hierarchies of the 4 Texture Zero Quark Mass Matrices and their
equal spacing rule}

\author{ F.  N.  Ndili \\
Physics Department \\
University of Houston, Houston, TX.77204, USA.}

\date{May, 2012}

\maketitle

\begin{abstract}
We show that the parameters of the quark mass matrices $M_u$ and
$M_d$ of a 4  texture zero model, exhibit an interesting
hierarchical \\ regularity we have called equal spacing rule, and
that this regularity leads us to a further observation that the
quark mass matrices $(M_u,M_d)$ appear to  exist in a plurality of
states, each state labelled by two numbers $(N, k)$. We give
interpretation to these observations.

\end{abstract}

E-mail: frank.ndili@gmail.com

\newpage

\section{Introduction}
The 4 texture zero hermitian quark mass matrix [1-12] continues to
present interesting study,  first with respect to the extent the
model reproduces experimental data, and secondly the extent any
such agreement with data is controlled by the hierarchies,
including geometric  hierarchy, of the various parameters of the
texture zero  quark mass matrices $M_u$ and $M_d$.  We focus
in the present article on the latter problem. \\

To reach the hierarchy problem we have first to state in some
detail the formalism of the 4 texture zero hermitian mass matrix.
This we do in section 2. In section 3 we adopt some approximations
that simplify our formalism, but enable us to calculate
analytically, the various parameters of our mass matrices $M_u$
and $M_d$. Armed with the parameter values, we proceed in section
4 to discuss the hierarchies of these parameters, and discover the
equal spacing rule that underlines the good agreement the 4
texture zero model has with experimental data. In section 5 we
consider the possible interpretation of the observed equal spacing
rule of the 4 texture zero mass matrices. We state our final
results  and conclusions in section 6. \\

\section{The 4 Texture zero quark mass matrix}
In their most familiar form, the mass matrices of the 4 texture
zero hermitian model are given by :
\begin{equation}\label{eq: ndili6}
 M^u =  \left(
\begin{array}{ccc}
0 & A_u & 0\\
A_u^*  & D_u & B_u\\
0 & B_u^* & C_u
\end{array}
\right)  ;  M^d =  \left(
\begin{array}{ccc}
0 & A_d & 0\\
A_d^*  & D_d & B_d\\
0 & B_d^* & C_d
\end{array}
\right)
\end{equation}
where $D_u, D_d$  are new non-zero elements in the (2,2) position,
not possessed by the original 6 texture zero quark mass matrix
model of Fritzsch [13,14]. The claim in the literature [1-12] is
that this 4 texture zero model endowed with the extra non-zero
(2,2) element, gives better agreement with data than the 6 texture
zero model of Fritzsch. Our interest is in seeing to what extent
this acclaimed success of the 4 texture zero model is tied up with
some intrinsic features, probably of hierarchical nature between
mass matrix elements, of the 4 texture zero model. \\

A standard formalism of analysis of $M_u$ and  $M_d$  is to first
separate out the complex phases of their elements:

$A_u = |A_u|e^{i\phi_{A_u}} , B_u = |B_u|e^{i\phi_{B_u}};$ \\

$A_d = |A_d|e^{i\phi_{A_d}} , B_d = |B_d|e^{i\phi_{B_d}},$ \\

where the phases are however unknown and need to be determined
later. Next one factors out these phases from $M^u$ and $M^d$, and
leaves real mass matrices $\hat M_u , \hat M_d$ defined by :
$M_{u} \rightarrow P_u \hat M_u P_u^\dagger$ and $M_{d}
\rightarrow P_d \hat M_d P_d^\dagger$, where
\begin{equation}\label{eq: ndili7}
 \hat M^u =  \left(
\begin{array}{ccc}
0 & |A_u| & 0\\
|A_u^*|  & D_u & |B_u |\\
0 & |B_u^*| & C_u
\end{array}
\right) ;  \hat M^d =  \left(
\begin{array}{ccc}
0 & |A_d| & 0\\
|A_d^*|  & D_d & |B_d|\\
0 & |B_d^*| & C_d
\end{array}
\right)
\end{equation}
and \\
\begin{equation}\label{eq: ndili8}
 P_u =  \left(
\begin{array}{ccc}
e^{i\alpha_u} & 0 & 0\\
0  & e^{i\beta_u} & 0 \\
0 & 0 & 1
\end{array} \right)  ;
P_d =  \left(
\begin{array}{ccc}
e^{i\alpha_d} & 0 & 0\\
0  & e^{i\beta_d} & 0 \\
0 & 0 & 1
\end{array} \right)
\end{equation}
with the relations:\\
$\phi_{A_u} = (\alpha_u - \beta_u ) ; \phi_{A_d} = (\alpha_d -
\beta_d );  \phi_{B_u} = \beta_u ; \phi_{B_d} = \beta_d$ so that
the product: $P =  P_u^\dagger P_d $
\begin{equation}\label{eq: ndili9}
 =  \left(
\begin{array}{ccc}
e^{i(\alpha_d-\alpha_u)} & 0 & 0\\
0  & e^{i(\beta_d - \beta_u)} & 0\\
0 & 0 & 1
\end{array}
\right) = \left(
\begin{array}{ccc}
e^{i((\phi_{A_d} - \phi_{A_u}) + (\phi_{B_d} -
\phi_{B_u}))} & 0 & 0\\
0  & e^{i(\phi_{B_d} - \phi_{B_u})} & 0\\
0 & 0 & 1
\end{array}
\right) =  \left(
\begin{array}{ccc}
e^{i\psi} & 0 & 0\\
0  & e^{i\phi} & 0\\
0 & 0 & 1
\end{array}
\right)
\end{equation}
where
\begin{equation}
 \psi = (\phi_1 + \phi_2) ; \phi_1 = (\phi_{A_d} -
\phi_{A_u}) ; \phi_2 = (\phi_{B_d} - \phi_{B_u}) = \phi.
\end{equation}

We next  diagonalize $\hat M_u$ and $\hat M_d$ through the
equations :  $\hat M_{u,d} \rightarrow O^T \hat M_{u,d} O = diag
(\lambda_1 , \lambda_2, \lambda_3)_{u,d},$ where O is some
orthogonal matrix whose elements we can calculate using the
standard method of eigenvectors. The $\lambda_1, \lambda_2,
\lambda_3$ are the eigenvalues we can identify with quark masses
in the following way: $(\lambda_1, \lambda_2, \lambda_3)_u = (m_u
, -m_c, m_t)$ and $(\lambda_1, \lambda_2, \lambda_3)_d = (m_d , -
m_s, m_b).$ The reason for the negative sign for $m_c, m_s$ will
be explained below.  \\

We invoke similarity transformation invariants to enable us relate
the unknown elements $|A|, |B|, D, C,$ of the mass matrices to the
mass eigenvalues $\lambda_i$. Because there are only three
invariant conditions while there are four unknown parameters in
this 4 texture zero model, we can remove at most only three of the
four parameters, leaving one unknown parameter we choose as C.
Then the above parameters are determined for now  as:
\begin{eqnarray}\label{eq: ndili10}
|A_u|  &=& \sqrt{\frac{-\lambda_1\lambda_2\lambda_3}{C_u}} =
\sqrt{\frac{m_u m_c m_t}{C_u}} \nonumber\\
|B_u|  &=&
\sqrt{\frac{(C_u-\lambda_1)(C_u-\lambda_2)(\lambda_3-C_u)}{C_u}}
=\sqrt{\frac{(C_u-m_u)(C_u+m_c)(m_t-C_u)}{C_u}} \nonumber\\
D_u  &=& \lambda_1 + \lambda_2 + \lambda_3 - C_u = m_u-m_c+m_t -
C_u \nonumber\\
|A_d|  &=& \sqrt{\frac{-\lambda_1\lambda_2\lambda_3}{C_d}} =
\sqrt{\frac{m_d m_s m_b}{C_d}} \nonumber\\
|B_d|  &=&
\sqrt{\frac{(C_d-\lambda_1)(C_d-\lambda_2)(\lambda_3-C_d)}{C_d}}
=\sqrt{\frac{(C_d-m_d)(C_d+m_s)(m_b-C_d)}{C_d}} \nonumber\\
D_d  &=& \lambda_1 + \lambda_2 + \lambda_3 - C_d = m_d-m_s+m_b -
C_d
\end{eqnarray}
We note the negative sign under the square root $\lambda$
expression for $|A_u|$ and $|A_d|$.  This requires that one of the
$\lambda_1, \lambda_2, \lambda_3$ eigenvalues must be negative so
that $|A|$ can be a real parameter like C. We can achieve this by
putting $(\lambda_1, \lambda_2, \lambda_3) = (m_1, -m_2 , m_3)$.
But other choices are possible, a fact Fritzsch and Xing [1] have
tried to deal with by introducing an $\eta = \pm 1$ sign
convention. We have not pursued that line of elaboration here. \\

Upon setting up the eigenvector equations of our new orthogonal
matrices, we determine their normalized eigenvectors to be :
\begin{equation}\label{eq: ndili11}
\vec X_1^u = \left(
\begin{array}{c}
x_1^u \\ y_1^u \\z_1^u
\end{array}
\right) =  \left(
\begin{array}{c}
\sqrt{\frac{\lambda_2\lambda_3(C_u-\lambda_1)}{C_u(\lambda_2-\lambda_1)(\lambda_3-\lambda_1)}}
\\ \\
\sqrt{\frac{\lambda_1(\lambda_1-C_u)}{(\lambda_2-\lambda_1)(\lambda_3-\lambda_1)}}\\
\\ \sqrt{\frac{\lambda_1(C_u-\lambda_2)(C_u-\lambda_3)}{C_u(\lambda_2-\lambda_1)(\lambda_3-\lambda_1)}}
\end{array}
\right) =  \left(
\begin{array}{c}
\sqrt{\frac{m_cm_t(C_u-m_u)}{C_u(m_c-m_u)(m_t-m_u)}}
\\ \\
\sqrt{\frac{m_u(C_u-m_u)}{(m_c+m_u)(m_t-m_u)}}\\
\\ \sqrt{\frac{m_u(C_u+ m_c)(m_t -C_u)}{C_u(m_c+m_u)(m_t-m_u)}}
\end{array}
\right)
\end{equation}

\begin{equation}\label{eq: ndili12}
\vec X_2^u = \left(
\begin{array}{c}
x_2^u \\ y_2^u \\z_2^u
\end{array}
\right) =  \left(
\begin{array}{c}
\sqrt{\frac{\lambda_1\lambda_3(\lambda_2-C_u)}{C_u(\lambda_2-\lambda_1)(\lambda_3-\lambda_2)}}
\\ \\
\sqrt{\frac{\lambda_2(C_u-\lambda_2)}{(\lambda_2-\lambda_1)(\lambda_3-\lambda_2)}}\\
\\ \sqrt{\frac{\lambda_2(C_u-\lambda_1)(\lambda_3-C_u)}{C_u(\lambda_2-\lambda_1)(\lambda_3-\lambda_2)}}
\end{array}
\right) = \left(
\begin{array}{c}
\sqrt{\frac{m_um_t(m_c+C_u)}{C_u(m_c+m_u)(m_t+m_c)}}
\\ \\
\sqrt{\frac{m_c(C_u+m_c)}{(m_c+m_u)(m_t+m_c)}}\\
\\ \sqrt{\frac{m_c(C_u-m_u)(m_t-C_u)}{C_u(m_c+m_u)(m_t+m_c)}}
\end{array}
\right)
\end{equation}

\begin{equation}\label{eq: ndili13}
\vec X_3^u = \left(
\begin{array}{c}
x_3^u \\ y_3^u \\z_3^u
\end{array}
\right) =  \left(
\begin{array}{c}
\sqrt{\frac{\lambda_1\lambda_2(C_u-\lambda_3)}{C_u(\lambda_3-\lambda_1)(\lambda_3-\lambda_2)}}\\
 \\ \sqrt{\frac{\lambda_3(\lambda_3-C_u)}{(\lambda_3-\lambda_1)(\lambda_3-\lambda_2)}}\\
\\ \sqrt{\frac{\lambda_3(C_u-\lambda_1)(C_u-\lambda_2)}{C_u(\lambda_3-\lambda_1)(\lambda_3-\lambda_2)}}
\end{array}
\right)   =  \left(
\begin{array}{c}
\sqrt{\frac{m_um_c(m_t-C_u)}{C_u(m_t-m_u)(m_t+m_c)}}\\
 \\ \sqrt{\frac{m_t(m_t-C_u)}{(m_t-m_u)(m_t + m_c)}}\\
\\ \sqrt{\frac{m_t(C_u-m_u)(C_u+m_c)}{C_u(m_t-m_u)(m_t +m_c)}}
\end{array}
\right)
\end{equation}
whence we determine $O_u$ as:
\begin{equation}\label{eq: ndili14}
O_u = \left(
\begin{array}{ccc}
x_1^u & x_2^u & x_3^u\\
y_1^u  & y_2^u & y_3^u \\
z_i^u & z_2^u & z_3^u
\end{array}
\right) = \left(
\begin{array}{ccc}
\sqrt{\frac{\lambda_2\lambda_3(C_u-\lambda_1)}{C_u(\lambda_2-\lambda_1)(\lambda_3-\lambda_1)}}
&
\sqrt{\frac{\lambda_1\lambda_3(\lambda_2-C_u)}{C_u(\lambda_2-\lambda_1)(\lambda_3-\lambda_2)}}
&
\sqrt{\frac{\lambda_1\lambda_2(C_u-\lambda_3)}{C_u(\lambda_3-\lambda_1)(\lambda_3-\lambda_2)}}\\
\\
\sqrt{\frac{\lambda_1(\lambda_1-C_u)}{(\lambda_2-\lambda_1)(\lambda_3-\lambda_1)}}
&
\sqrt{\frac{\lambda_2(C_u-\lambda_2)}{(\lambda_2-\lambda_1)(\lambda_3-\lambda_2)}}
&
\sqrt{\frac{\lambda_3(\lambda_3-C_u)}{(\lambda_3-\lambda_1)(\lambda_3-\lambda_2)}}\\
\\
 \sqrt{\frac{\lambda_1(C_u-\lambda_2)(C_u-\lambda_3)}{C_u(\lambda_2-\lambda_1)(\lambda_3-\lambda_1)}} &
 \sqrt{\frac{\lambda_2(C_u-\lambda_1)(\lambda_3-C_u)}{C_u(\lambda_2-\lambda_1)(\lambda_3-\lambda_2)}} &
\sqrt{\frac{\lambda_3(C_u-\lambda_1)(C_u-\lambda_2)}{C_u(\lambda_3-\lambda_1)(\lambda_3-\lambda_2)}}
\end{array} \right)
\end{equation}
or \\
\begin{equation}\label{eq: ndili15}
O_u = \left(
\begin{array}{ccc}
x_1^u & x_2^u & x_3^u\\
y_1^u  & y_2^u & y_3^u \\
z_i^u & z_2^u & z_3^u
\end{array}
\right) = \left(
\begin{array}{ccc}
\sqrt{\frac{m_cm_t(C_u-m_u)}{C_u(m_c-m_u)(m_t-m_u)}} &
\sqrt{\frac{m_um_t(m_c+C_u)}{C_u(m_c+m_u)(m_t+m_c)}} &
\sqrt{\frac{m_um_c(m_t-C_u)}{C_u(m_t-m_u)(m_t+m_c)}}\\ \\
\sqrt{\frac{m_u(C_u-m_u)}{(m_c+m_u)(m_t-m_u)}} &
\sqrt{\frac{m_c(C_u+m_c)}{(m_c+m_u)(m_t+m_c)}} &
\sqrt{\frac{m_t(m_t-C_u)}{(m_t-m_u)(m_t + m_c)}}\\ \\
\sqrt{\frac{m_u(C_u+ m_c)(m_t -C_u)}{C_u(m_c+m_u)(m_t-m_u)}} &
\sqrt{\frac{m_c(C_u-m_u)(m_t-C_u)}{C_u(m_c+m_u)(m_t+m_c)}} &
\sqrt{\frac{m_t(C_u-m_u)(C_u+m_c)}{C_u(m_t-m_u)(m_t +m_c)}}
\end{array} \right)
\end{equation}

Similarly we have for the $dsb$ sector:
\begin{equation}\label{eq: ndili16}
\vec X_1^d = \left(
\begin{array}{c}
x_1^d \\ y_1^d \\z_1^d
\end{array}
\right) =  \left(
\begin{array}{c}
\sqrt{\frac{\lambda_2\lambda_3(C_d-\lambda_1)}{C_d(\lambda_2-\lambda_1)(\lambda_3-\lambda_1)}}
\\ \\ \sqrt{\frac{\lambda_1(\lambda_1-C_d)}{(\lambda_2-\lambda_1)(\lambda_3-\lambda_1)}}
\\ \\ \sqrt{\frac{\lambda_1(C_d-\lambda_2)(C_d-\lambda_3)}{C_d(\lambda_2-\lambda_1)(\lambda_3-\lambda_1)}}
\end{array}
\right) =  \left(
\begin{array}{c}
\sqrt{\frac{m_sm_b(C_d-m_d)}{C_d(m_s-m_d)(m_b-m_d)}}
\\ \\
\sqrt{\frac{m_d(C_d-m_d)}{(m_s+m_d)(m_b-m_d)}}\\
\\ \sqrt{\frac{m_d(C_d+ m_s)(m_b -C_d)}{C_d(m_s+m_d)(m_b-m_d)}}
\end{array}
\right)
\end{equation}

\begin{equation}\label{eq: ndili17}
\vec X_2^d = \left(
\begin{array}{c}
x_2^d \\ y_2^d \\z_2^d
\end{array}
\right) =  \left(
\begin{array}{c}
\sqrt{\frac{\lambda_1\lambda_3(\lambda_2-C_d)}{C_d(\lambda_2-\lambda_1)(\lambda_3-\lambda_2)}}
\\ \\ \sqrt{\frac{\lambda_2(C_d-\lambda_2)}{(\lambda_2-\lambda_1)(\lambda_3-\lambda_2)}}
\\ \\ \sqrt{\frac{\lambda_2(C_d-\lambda_1)(\lambda_3-C_d)}{C_d(\lambda_2-\lambda_1)(\lambda_3-\lambda_2)}}
\end{array}
\right) = \left(
\begin{array}{c}
\sqrt{\frac{m_dm_b(m_s+C_d)}{C_d(m_s+m_d)(m_b+m_s)}}
\\ \\
\sqrt{\frac{m_s(C_d+m_s)}{(m_s+m_d)(m_b+m_s)}}\\
\\ \sqrt{\frac{m_s(C_d-m_d)(m_b-C_d)}{C_d(m_s+m_d)(m_b+m_s)}}
\end{array}
\right)
\end{equation}

\begin{equation}\label{eq: ndili18}
\vec X_3^d = \left(
\begin{array}{c}
x_3^d \\ y_3^d \\z_3^d
\end{array}
\right) =  \left(
\begin{array}{c}
\sqrt{\frac{\lambda_1\lambda_2(C_d-\lambda_3)}{C_d(\lambda_3-\lambda_1)(\lambda_3-\lambda_2)}}\\
 \\ \sqrt{\frac{\lambda_3(\lambda_3-C_d)}{(\lambda_3-\lambda_1)(\lambda_3-\lambda_2)}}\\
\\ \sqrt{\frac{\lambda_3(C_d-\lambda_1)(C_d-\lambda_2)}{C_d(\lambda_3-\lambda_1)(\lambda_3-\lambda_2)}}
\end{array}
\right)  =  \left(
\begin{array}{c}
\sqrt{\frac{m_dm_s(m_b-C_d)}{C_d(m_b-m_d)(m_b+m_s)}}\\
 \\ \sqrt{\frac{m_b(m_b-C_d)}{(m_b-m_d)(m_b + m_s)}}\\
\\ \sqrt{\frac{m_b(C_d-m_d)(C_d+m_s)}{C_d(m_b-m_d)(m_b +m_s)}}
\end{array}
\right)
\end{equation}
whence we determine $O_d$ as:
\begin{equation}\label{eq: ndili19}
O_d = \left(
\begin{array}{ccc}
x_1^d & x_2^d & x_3^d\\
y_1^d  & y_2^d & y_3^d \\
z_i^d & z_2^d & z_3^d
\end{array}
\right) = \left(
\begin{array}{ccc}
\sqrt{\frac{\lambda_2\lambda_3(C_d-\lambda_1)}{C_d(\lambda_2-\lambda_1)(\lambda_3-\lambda_1)}}
&
\sqrt{\frac{\lambda_1\lambda_3(\lambda_2-C_d)}{C_d(\lambda_2-\lambda_1)(\lambda_3-\lambda_2)}}
&
\sqrt{\frac{\lambda_1\lambda_2(C_d-\lambda_3)}{C_d(\lambda_3-\lambda_1)(\lambda_3-\lambda_2)}}\\
\\
\sqrt{\frac{\lambda_1(\lambda_1-C_d)}{(\lambda_2-\lambda_1)(\lambda_3-\lambda_1)}}
&
\sqrt{\frac{\lambda_2(C_d-\lambda_2)}{(\lambda_2-\lambda_1)(\lambda_3-\lambda_2)}}
&
\sqrt{\frac{\lambda_3(\lambda_3-C_d)}{(\lambda_3-\lambda_1)(\lambda_3-\lambda_2)}}\\
\\
 \sqrt{\frac{\lambda_1(C_d-\lambda_2)(C_d-\lambda_3)}{C_d(\lambda_2-\lambda_1)(\lambda_3-\lambda_1)}} &
 \sqrt{\frac{\lambda_2(C_d-\lambda_1)(\lambda_3-C_d)}{C_d(\lambda_2-\lambda_1)(\lambda_3-\lambda_2)}} &
\sqrt{\frac{\lambda_3(C_d-\lambda_1)(C_d-\lambda_2)}{C_d(\lambda_3-\lambda_1)(\lambda_3-\lambda_2)}}
\end{array} \right)
\end{equation}
or \\
\begin{equation}\label{eq: ndili20}
O_d = \left(
\begin{array}{ccc}
x_1^d & x_2^d & x_3^d\\
y_1^d  & y_2^d & y_3^d \\
z_i^d & z_2^d & z_3^d
\end{array}
\right) =  \left(
\begin{array}{ccc}
\sqrt{\frac{m_sm_b(C_d-m_d)}{C_d(m_s-m_d)(m_b-m_d)}} &
\sqrt{\frac{m_dm_b(m_s+C_d)}{C_d(m_s+m_d)(m_b+m_s)}} &
\sqrt{\frac{m_dm_s(m_b-C_d)}{C_d(m_b-m_d)(m_b+m_s)}}\\ \\
\sqrt{\frac{m_d(C_d-m_d)}{(m_s+m_d)(m_b-m_d)}} &
\sqrt{\frac{m_s(C_d+m_s)}{(m_s+m_d)(m_b+m_s)}} &
\sqrt{\frac{m_b(m_b-C_d)}{(m_b-m_d)(m_b + m_s)}}\\ \\
\sqrt{\frac{m_d(C_d+ m_s)(m_b -C_d)}{C_d(m_s+m_d)(m_b-m_d)}} &
\sqrt{\frac{m_s(C_d-m_d)(m_b-C_d)}{C_d(m_s+m_d)(m_b+m_s)}} &
\sqrt{\frac{m_b(C_d-m_d)(C_d+m_s)}{C_d(m_b-m_d)(m_b +m_s)}}
\end{array} \right)
\end{equation}

Given now $O_u$ and $O_d$, we compute elements of $V_{CKM}$
predicted by the four texture zero model.  We have :
\begin{equation}\label{eq: ndili21}
V_{CKM} = {O'}_u^\dagger {O'}_d = O_u^\dagger P_u^\dagger P_d O_d
= O_u^\dagger P O_d
\end{equation}
where $P =  P_u^\dagger P_d $ was given earlier in equation (4).

The various elements of $V_{CKM}$ are thereafter given by :
\begin{eqnarray}\label{eq: ndili22}
V_{ud}&=& x_1^u x_1^d e^{i\psi}+y_1^u y_1^d e^{i\phi}+z_1^u z_1^d \nonumber\\
V_{us}&=& x_1^u x_2^d e^{i\psi}+y_1^u y_2^d e^{i\phi}+z_1^u z_2^d \nonumber\\
V_{ub}&=& x_1^u x_3^d e^{i\psi}+y_1^u y_3^d e^{i\phi}+z_1^u z_3^d \nonumber\\
V_{cd}&=& x_2^u x_1^d e^{i\psi}+y_2^u y_1^d e^{i\phi}+z_2^u z_1^d \nonumber\\
V_{cs}&=& x_2^u x_2^d e^{i\psi}+y_2^u y_2^d e^{i\phi}+z_2^u z_2^d \nonumber\\
V_{cb}&=& x_2^u x_3^d e^{i\psi}+y_2^u y_3^d e^{i\phi}+z_2^u z_3^d \nonumber\\
V_{td}&=& x_3^u x_1^d e^{i\psi}+y_3^u y_1^d e^{i\phi}+z_3^u z_1^d \nonumber\\
V_{ts}&=& x_3^u x_2^d e^{i\psi}+y_3^u y_2^d e^{i\phi}+z_3^u z_2^d \nonumber\\
V_{tb}&=& x_3^u x_3^d e^{i\psi}+y_3^u y_3^d e^{i\phi}+z_3^u z_3^d
\end{eqnarray}
Explicitly we obtain:
\begin{eqnarray}\label{eq: ndili23}
V_{ud}&=& e^{i
\psi}\sqrt{\frac{m_cm_t(C_u-m_u)}{C_u(m_c-m_u)(m_t-m_u)}}
\sqrt{\frac{m_sm_b(C_d-m_d)}{C_d(m_s-m_d)(m_b-m_d)}} \nonumber\\
 & & {} + e^{i\phi}\sqrt{\frac{m_u(C_u-m_u)}{(m_c+m_u)(m_t-m_u)}}
\sqrt{\frac{m_d(C_d-m_d)}{(m_s+m_d)(m_b-m_d)}} \nonumber\\
& & {} +
\sqrt{\frac{m_u(C_u+m_c)(m_t-C_u)}{C_u(m_c+m_u)(m_t-m_u)}}
 \sqrt{\frac{m_d(C_d+ m_s)(m_b -C_d)}{C_d(m_s+m_d)(m_b-m_d)}}
\end{eqnarray}

\begin{eqnarray}\label{eq: ndili24}
V_{us}&=& e^{i
\psi}\sqrt{\frac{m_cm_t(C_u-m_u)}{C_u(m_c-m_u)(m_t-m_u)}}
\sqrt{\frac{m_dm_b(m_s+C_d)}{C_d(m_s+m_d)(m_b+m_s)}} \nonumber\\
  & & {} + e^{i\phi}\sqrt{\frac{m_u(C_u-m_u)}{(m_c+m_u)(m_t-m_u)}}
\sqrt{\frac{m_s(C_d+m_s)}{(m_s+m_d)(m_b+m_s)}} \nonumber\\
 & & {} + \sqrt{\frac{m_u(C_u+ m_c)(m_t -C_u)}{C_u(m_c+m_u)(m_t-m_u)}}
\sqrt{\frac{m_s(C_d-m_d)(m_b-C_d)}{C_d(m_s+m_d)(m_b+m_s)}}
\end{eqnarray}

\begin{eqnarray}\label{eq: ndili25}
V_{ub}&=& e^{i
\psi}\sqrt{\frac{m_cm_t(C_u-m_u)}{C_u(m_c-m_u)(m_t-m_u)}}
\sqrt{\frac{m_dm_s(m_b-C_d)}{C_d(m_b-m_d)(m_b+m_s)}} \nonumber\\
 & & {} + e^{i\phi}\sqrt{\frac{m_u(C_u-m_u)}{(m_c+m_u)(m_t-m_u)}}
\sqrt{\frac{m_b(m_b-C_d)}{(m_b-m_d)(m_b + m_s)}} \nonumber\\
 & & {} + \sqrt{\frac{m_u(C_u+ m_c)(m_t -C_u)}{C_u(m_c+m_u)(m_t-m_u)}}
\sqrt{\frac{m_b(C_d-m_d)(C_d+m_s)}{C_d(m_b-m_d)(m_b +m_s)}}
\end{eqnarray}

\begin{eqnarray}\label{eq: ndili26}
V_{cd}&=& e^{i\psi}
\sqrt{\frac{m_um_t(m_c+C_u)}{C_u(m_c+m_u)(m_t+m_c)}}
\sqrt{\frac{m_sm_b(C_d-m_d)}{C_d(m_s-m_d)(m_b-m_d)}} \nonumber\\
 & & {} + e^{i\phi}
 \sqrt{\frac{m_c(C_u+m_c)}{(m_c+m_u)(m_t+m_c)}}
\sqrt{\frac{m_d(C_d-m_d)}{(m_s+m_d)(m_b-m_d)}} \nonumber\\
 & & {} + \sqrt{\frac{m_c(C_u-m_u)(m_t-C_u)}{C_u(m_c+m_u)(m_t+m_c)}}
\sqrt{\frac{m_d(C_d+ m_s)(m_b -C_d)}{C_d(m_s+m_d)(m_b-m_d)}}
\end{eqnarray}

\begin{eqnarray}\label{eq: ndili27}
V_{cs}&=& e^{i\psi}
\sqrt{\frac{m_um_t(m_c+C_u)}{C_u(m_c+m_u)(m_t+m_c)}}
\sqrt{\frac{m_dm_b(m_s+C_d)}{C_d(m_s+m_d)(m_b+m_s)}} \nonumber\\
 & & {} + e^{i\phi}
\sqrt{\frac{m_c(C_u+m_c)}{(m_c+m_u)(m_t+m_c)}}
\sqrt{\frac{m_s(C_d+m_s)}{(m_s+m_d)(m_b+m_s)}} \nonumber\\
 & & {} + \sqrt{\frac{m_c(C_u-m_u)(m_t-C_u)}{C_u(m_c+m_u)(m_t+m_c)}}
\sqrt{\frac{m_s(C_d-m_d)(m_b-C_d)}{C_d(m_s+m_d)(m_b+m_s)}}
\end{eqnarray}

\begin{eqnarray}\label{eq: ndili28}
V_{cb}&=& e^{i\psi}
\sqrt{\frac{m_um_t(m_c+C_u)}{C_u(m_c+m_u)(m_t+m_c)}}
\sqrt{\frac{m_dm_s(m_b-C_d)}{C_d(m_b-m_d)(m_b+m_s)}} \nonumber\\
 & & {} + e^{i\phi}
\sqrt{\frac{m_c(C_u+m_c)}{(m_c+m_u)(m_t+m_c)}}
\sqrt{\frac{m_b(m_b-C_d)}{(m_b-m_d)(m_b + m_s)}} \nonumber\\
 & & {} + \sqrt{\frac{m_c(C_u-m_u)(m_t-C_u)}{C_u(m_c+m_u)(m_t+m_c)}}
\sqrt{\frac{m_b(C_d-m_d)(C_d+m_s)}{C_d(m_b-m_d)(m_b +m_s)}}
\end{eqnarray}

\begin{eqnarray}\label{eq: ndili29}
V_{td}&=& e^{i\psi}
\sqrt{\frac{m_um_c(m_t-C_u)}{C_u(m_t-m_u)(m_t+m_c)}}
\sqrt{\frac{m_sm_b(C_d-m_d)}{C_d(m_s-m_d)(m_b-m_d)}} \nonumber\\
 & & {} + e^{i\phi}
\sqrt{\frac{m_t(m_t-C_u)}{(m_t-m_u)(m_t + m_c)}}
\sqrt{\frac{m_d(C_d-m_d)}{(m_s+m_d)(m_b-m_d)}} \nonumber\\
 & & {} + \sqrt{\frac{m_t(C_u-m_u)(C_u+m_c)}{C_u(m_t-m_u)(m_t +m_c)}}
\sqrt{\frac{m_d(C_d+ m_s)(m_b -C_d)}{C_d(m_s+m_d)(m_b-m_d)}}
\end{eqnarray}

\begin{eqnarray}\label{eq: ndili30}
V_{ts}&=& e^{i\psi}
\sqrt{\frac{m_um_c(m_t-C_u)}{C_u(m_t-m_u)(m_t+m_c)}}
\sqrt{\frac{m_dm_b(m_s+C_d)}{C_d(m_s+m_d)(m_b+m_s)}} \nonumber\\
 & & {} + e^{i\phi}
\sqrt{\frac{m_t(m_t-C_u)}{(m_t-m_u)(m_t + m_c)}}
\sqrt{\frac{m_s(C_d+m_s)}{(m_s+m_d)(m_b+m_s)}} \nonumber\\
 & & {} + \sqrt{\frac{m_t(C_u-m_u)(C_u+m_c)}{C_u(m_t-m_u)(m_t +m_c)}}
\sqrt{\frac{m_s(C_d-m_d)(m_b-C_d)}{C_d(m_s+m_d)(m_b+m_s)}}
\end{eqnarray}

\begin{eqnarray}\label{eq: ndili31}
V_{tb}&=& e^{i\psi}
\sqrt{\frac{m_um_c(m_t-C_u)}{C_u(m_t-m_u)(m_t+m_c)}}
\sqrt{\frac{m_b(C_d-m_d)(C_d+m_s)}{C_d(m_b-m_d)(m_b +m_s)}} \nonumber\\
 & & {} + e^{i\phi}
\sqrt{\frac{m_t(m_t-C_u)}{(m_t-m_u)(m_t + m_c)}}
\sqrt{\frac{m_b(m_b-C_d)}{(m_b-m_d)(m_b + m_s)}} \nonumber\\
 & & {} + \sqrt{\frac{m_t(C_u-m_u)(C_u+m_c)}{C_u(m_t-m_u)(m_t +m_c)}}
\sqrt{\frac{m_b(C_d-m_d)(C_d+m_s)}{C_d(m_b-m_d)(m_b +m_s)}}
\end{eqnarray}

We note that unlike the 6 texture zero case where each element
$|V_{ij}|$ is expressed entirely in terms of the known quark
masses (and two phase angles), in the present case of 4 texture
zero the unknown parameters $C_u$ and $C_d$  appear in every
element $|V_{ij}|$, plus the same two phase angles $\psi $ and
$\phi$. These parameters can play the role of adjustable
parameters needed to accommodate a range of measured flavor
mixing. More generally the 4 texture zero model with the above
feature allows us to regard $C_u, C_d$ as well as all the other
parameters $|A|, |B|. D,$ linked to $C_u, C_d$ by equation (6), as
adjustable parameters, a feature the 6 texture zero model did not
have. Armed with this feature we can analyze the 4 texture zero
model further and determine directly the magnitudes and
hierarchies of these parameters, supported by experimental data.
\\

\section{Approximations and numerical Analysis of the 4 texture zero model}
We adopt at this stage, a minimal set of reasonable assumptions
designed to simplify the various $|V_{ij}|$ expressions above. We
invoke the known hierarchy of quark masses: $m_t \gg m_c \gg m_u ;
m_b \gg m_s \gg m_d$ ; also we assume :$C_u \gg m_c \gg m_u ; C_d
\gg m_s \gg m_d$, but leave the ratios $C_u/m_t , C_d/m_b$ as
quantities we need to specifically calculate and determine. With
these approximations the above $V_{ij}$ quantities become:
\begin{eqnarray}\label{eq: ndili32}
V_{ud}& \approx & e^{i \psi} +
e^{i\phi}\sqrt{\frac{m_um_d}{m_cm_s}}\sqrt{\frac{C_u}{m_t}}\sqrt{\frac{C_d}{m_b}}
+ \sqrt{\frac{m_um_d}{m_cm_s}}\sqrt{(1 - \frac{C_u}{m_t})}\sqrt{(1
- \frac{C_d}{m_b})}
\end{eqnarray}
or
\begin{eqnarray}\label{eq: ndili33}
|V_{ud}| & \approx & \left| 1 +
e^{-i\phi_1}\sqrt{\frac{m_um_d}{m_cm_s}}\sqrt{\frac{C_u}{m_t}}\sqrt{\frac{C_d}{m_b}}
+ e^{-i\psi}\sqrt{\frac{m_um_d}{m_cm_s}}\sqrt{(1 -
\frac{C_u}{m_t})}\sqrt{(1 - \frac{C_d}{m_b})}\right| \nonumber\\
  & \approx & 1
\end{eqnarray}

\begin{eqnarray}\label{eq: ndili34}
V_{us}& \approx & e^{i\psi}\sqrt{\frac{m_d}{m_s}} +
e^{i\phi}\sqrt{\frac{m_u}{m_c}}\sqrt{\frac{C_u}{m_t}}\sqrt{\frac{C_d}{m_b}}
+ \sqrt{\frac{m_u}{m_c}}\sqrt{(1 - \frac{C_u}{m_t})}\sqrt{(1 -
\frac{C_d}{m_b})} \nonumber\\
  &  &
\end{eqnarray}
or
\begin{eqnarray}\label{eq: ndili35}
|V_{us}| & \approx & \left|\sqrt{\frac{m_d}{m_s}} +
e^{-i\phi_1}\sqrt{\frac{m_u}{m_c}}\sqrt{\frac{C_u}{m_t}}\sqrt{\frac{C_d}{m_b}}
+ e^{-i\psi}\sqrt{\frac{m_u}{m_c}}\sqrt{(1 -
\frac{C_u}{m_t})}\sqrt{(1 -
\frac{C_d}{m_b})} \right| \nonumber\\
  &  &
\end{eqnarray}

\begin{eqnarray}\label{eq: ndili36}
V_{ub}&\approx & e^{i
\psi}\sqrt{\frac{m_dm_s}{m_bC_d}(1-\frac{C_d}{m_b})} +
e^{i\phi}\sqrt{\frac{m_u}{m_c}}\sqrt{\frac{C_u}{m_t}(1-\frac{C_d}{m_b})}
+ \sqrt{\frac{m_u}{m_c}}\sqrt{\frac{C_d}{m_b}(1- \frac{C_u}{m_t})}
\nonumber\\
  &  &
\end{eqnarray}
or
\begin{eqnarray}\label{eq: ndili37}
|V_{ub}| & \approx & \left|
\sqrt{\frac{m_dm_s}{m_bC_d}(1-\frac{C_d}{m_b})} +
e^{-i\phi_1}\sqrt{\frac{m_u}{m_c}}\sqrt{\frac{C_u}{m_t}(1-\frac{C_d}{m_b})}
+ e^{-i\psi} \sqrt{\frac{m_u}{m_c}}\sqrt{\frac{C_d}{m_b}(1-
\frac{C_u}{m_t})} \right| \nonumber\\
  & \approx & \left| \sqrt{\frac{m_u}{m_c}}\sqrt{\frac{C_u}{m_t}}\sqrt{(1-\frac{C_d}{m_b})}
+ e^{-i\phi_2}
\sqrt{\frac{m_u}{m_c}}\sqrt{\frac{C_d}{m_b}}\sqrt{(1-
\frac{C_u}{m_t})} \right|
\end{eqnarray}
\begin{eqnarray}\label{eq: ndili38}
V_{cd}& \approx & e^{i\psi} \sqrt{\frac{m_u}{m_c}} +
e^{i\phi}\sqrt{\frac{m_d}{m_s}}\sqrt{\frac{C_u}{m_t}}\sqrt{\frac{C_d}{m_b}}
+ \sqrt{\frac{m_d}{m_s}(1
-\frac{C_d}{m_b})}\sqrt{(1-\frac{C_u}{m_t})} \nonumber\\
   &  &
\end{eqnarray}
or
\begin{eqnarray}\label{eq: ndili39}
|V_{cd}| & \approx & \left|\sqrt{\frac{m_u}{m_c}} +
e^{-i\phi_1}\sqrt{\frac{m_d}{m_s}}\sqrt{\frac{C_u}{m_t}}\sqrt{\frac{C_d}{m_b}}
+ e^{-i\psi}\sqrt{\frac{m_d}{m_s}}\sqrt{(1
-\frac{C_d}{m_b})}\sqrt{(1-\frac{C_u}{m_t})}\right| \nonumber\\
   &  &
\end{eqnarray}
\begin{eqnarray}\label{eq: ndili40}
V_{cs}& \approx & e^{i\psi}
\sqrt{\frac{m_u}{m_c}}\sqrt{\frac{m_d}{m_s}} +
e^{i\phi}\sqrt{\frac{C_u}{m_t}}\sqrt{\frac{C_d}{m_b}} +
\sqrt{(1-\frac{C_u}{m_t})}\sqrt{(1-\frac{C_d}{m_b})} \nonumber\\
  &  &
\end{eqnarray}
or
\begin{eqnarray}\label{eq: ndili41}
|V_{cs}| & \approx & \left|
\sqrt{\frac{C_u}{m_t}}\sqrt{\frac{C_d}{m_b}} + e^{i\phi_1}
\sqrt{\frac{m_u}{m_c}}\sqrt{\frac{m_d}{m_s}} + e^{-i\phi}
\sqrt{(1-\frac{C_u}{m_t})}\sqrt{(1-\frac{C_d}{m_b})} \right| \nonumber\\
  & \approx & \sqrt{\frac{C_u}{m_t}}\sqrt{\frac{C_d}{m_b}}
\end{eqnarray}
\begin{eqnarray}\label{eq: ndili42}
V_{cb}& \approx & e^{i\psi}
\sqrt{\frac{m_u}{m_c}}\sqrt{\frac{m_dm_s}{m_bC_d}} +
e^{i\phi}\sqrt{\frac{C_u}{m_t}}\sqrt{(1 - \frac{C_d}{m_b})} +
\sqrt{\frac{C_d}{m_b}}\sqrt{(1-\frac{C_u}{m_t})} \nonumber\\
 &  &
\end{eqnarray}
or
\begin{eqnarray}\label{eq: ndili43}
|V_{cb}| & \approx & \left|
\sqrt{\frac{m_u}{m_c}}\sqrt{\frac{m_dm_s}{m_bC_d}} +
e^{-i\phi_1}\sqrt{\frac{C_u}{m_t}}\sqrt{(1 - \frac{C_d}{m_b})} +
e^{-i\psi}\sqrt{\frac{C_d}{m_b}}\sqrt{(1-\frac{C_u}{m_t})} \right| \nonumber\\
 & \approx & \left|\sqrt{\frac{C_u}{m_t}}\sqrt{(1 - \frac{C_d}{m_b})} +
e^{-i\phi_2}\sqrt{\frac{C_d}{m_b}}\sqrt{(1-\frac{C_u}{m_t})}\right|
\end{eqnarray}
\begin{eqnarray}\label{eq: ndili44}
V_{td}& \approx & e^{i\psi} \sqrt{\frac{m_um_c}{m_tC_u}}\sqrt{(1-
\frac{C_u}{m_t})} + e^{i\phi}
\sqrt{\frac{m_d}{m_s}}\sqrt{\frac{C_d}{m_b}}\sqrt{(1-
\frac{C_u}{m_t})} +
\sqrt{\frac{C_u}{m_t}}\sqrt{\frac{m_d}{m_s}}\sqrt{(1-\frac{C_d}{m_b})}
\nonumber\\
  &  &
\end{eqnarray}
or
\begin{eqnarray}\label{eq: ndili45}
|V_{td}| & \approx & \left| \sqrt{\frac{m_um_c}{m_tC_u}}\sqrt{(1-
\frac{C_u}{m_t})} + e^{-i\phi_1}
\sqrt{\frac{m_d}{m_s}}\sqrt{\frac{C_d}{m_b}}\sqrt{(1-
\frac{C_u}{m_t})} + e^{-i\psi}
\sqrt{\frac{C_u}{m_t}}\sqrt{\frac{m_d}{m_s}}\sqrt{(1-\frac{C_d}{m_b})}
\right| \nonumber\\
  & \approx   & \left| \sqrt{\frac{m_d}{m_s}}\sqrt{\frac{C_d}{m_b}}\sqrt{(1-
\frac{C_u}{m_t})} + e^{-i\phi_2}
\sqrt{\frac{C_u}{m_t}}\sqrt{\frac{m_d}{m_s}}\sqrt{(1-\frac{C_d}{m_b})}
\right|
\end{eqnarray}
\begin{eqnarray}\label{eq: ndili46}
V_{ts}& \approx & e^{i\psi}
\sqrt{\frac{m_d}{m_s}}\sqrt{\frac{m_um_c}{m_tC_u}}\sqrt{(1 -
\frac{C_u}{m_t})} +  e^{i\phi}\sqrt{\frac{C_d}{m_b}}\sqrt{(1 -
\frac{C_u}{m_t})} +  \sqrt{\frac{C_u}{m_t}}\sqrt{(1 -
\frac{C_d}{m_b})} \nonumber\\
   &   &
\end{eqnarray}
or
\begin{eqnarray}\label{eq: ndili47}
|V_{ts}| & \approx & \left|
\sqrt{\frac{m_d}{m_s}}\sqrt{\frac{m_um_c}{m_tC_u}}\sqrt{(1 -
\frac{C_u}{m_t})} +  e^{-i\phi_1}\sqrt{\frac{C_d}{m_b}}\sqrt{(1 -
\frac{C_u}{m_t})} + e^{-i\psi} \sqrt{\frac{C_u}{m_t}}\sqrt{(1 -
\frac{C_d}{m_b})} \right| \nonumber\\
   & \approx  & \left| \sqrt{\frac{C_d}{m_b}}\sqrt{(1 -
\frac{C_u}{m_t})} + e^{-i\phi_2} \sqrt{\frac{C_u}{m_t}}\sqrt{(1 -
\frac{C_d}{m_b})} \right|
\end{eqnarray}
\begin{eqnarray}\label{eq: ndili48}
V_{tb}& \approx & e^{i\psi}
\sqrt{\frac{C_d}{m_b}}\sqrt{\frac{m_um_c}{m_tC_u}}\sqrt{(1 -
\frac{C_u}{m_t})} + e^{i\phi}\sqrt{(1 - \frac{C_u}{m_t})}\sqrt{(1
- \frac{C_d}{m_b})} + \sqrt{\frac{C_u}{m_t}}\sqrt{\frac{C_d}{m_b}}
\nonumber\\
   &  &
\end{eqnarray}
or
\begin{eqnarray}\label{eq: ndili49}
|V_{tb}| & \approx & \left|
\sqrt{\frac{C_u}{m_t}}\sqrt{\frac{C_d}{m_b}} +  e^{i\psi}
\sqrt{\frac{C_d}{m_b}}\sqrt{\frac{m_um_c}{m_tC_u}}\sqrt{(1 -
\frac{C_u}{m_t})} + e^{i\phi}\sqrt{(1 - \frac{C_u}{m_t})}\sqrt{(1
- \frac{C_d}{m_b})} \right|  \nonumber\\
   & \approx  & \sqrt{\frac{C_u}{m_t}}\sqrt{\frac{C_d}{m_b}}
\end{eqnarray}

Notably, we see in this 4 texture zero model, that many of the
$|V_{ij}|$ quantities  owe any non-zero value they may have
experimentally to  the values we attach to the ratios  $C_u/m_t$
and $C_d/m_b$, and by implication from equation(6), to the
hierarchical values of the rest parameters of $M_u$ and $M_d$.
Such critical dependence of the predictions of a texture zero
model on hierarchies of the parameters of the model  while present
in the 4 texture zero case as shown above, are not present in the
6 texture zero model. We show this helps explain and illuminate
some of the lack of consistency and agreement of the 6 texture
zero model with experiment. \\

Thus we note from our equation (31) that the familiar result for
the Cabibbo angle obtained in the 6 texture zero model [13, 14]
namely:
\begin{equation}\label{eq: ndili3A}
 |V_{us}| = |\sin \theta_C| \approx \left| \sqrt{\frac{m_d}{m_s}} +
 e^{-i\phi_1} \sqrt{\frac{m_u}{m_c}} \right| \approx |V_{cd}|
\end{equation}

is only reproduced here (in the 4 texture zero formalism), when we
assume $\sqrt{C_u/m_t} = \sqrt{C_d/m_b} = 1 $ and $\sqrt{1 -
C_u/m_t} = \sqrt{1 - C_d/m_b} =0$. But then from our equations
(33), (39), (41), (43), the same assumption leads to zero values
for $|V_{ub}|, |V_{cb}|, |V_{td}|$ and $V_{ts}|$. We see this as a
good reason why the 6 texture zero model has difficulty with its
other predicted relations, namely:
\begin{equation}\label{eq: ndili4A}
\frac{|V_{ub}|}{|V_{cb}|} = \sqrt{\frac{m_u}{m_c}}
\end{equation}
and
\begin{equation}\label{eq: ndili5A}
\frac{|V_{td}|}{|V_{ts}|} = \sqrt{\frac{m_d}{m_s}}
\end{equation}
Equations (47) and (48) would appear not intrinsically compatible
with equation (46). The only way equations (47) and (48) can arise
from our 4 texture zero model, is if we adopt a different
approximation for $V_{ub}$ and $V_{cb}$ namely :
\begin{eqnarray}\label{eq: ndili50C}
V_{ub}&\approx & \sqrt{\frac{m_u}{m_c}}\left[
e^{i\phi}\sqrt{\frac{C_u}{m_t}}\sqrt{(1-\frac{C_d}{m_b})} +
\sqrt{\frac{C_d}{m_b}}\sqrt{(1- \frac{C_u}{m_t})}\right]
\nonumber\\
  &  &
\end{eqnarray}
and
\begin{eqnarray}\label{eq: ndili50}
V_{cb}& \approx & e^{i\phi}\sqrt{\frac{C_u}{m_t}}\sqrt{(1 -
\frac{C_d}{m_b})} +
\sqrt{\frac{C_d}{m_b}}\sqrt{(1-\frac{C_u}{m_t})} \nonumber\\
 &  &
\end{eqnarray}
 whence
\begin{equation}
\frac{|V_{ub}|}{|V_{cb}|} = \sqrt{\frac{m_u}{m_c}}
\end{equation}
as in the 6 texture zero model. \\

By similarly dropping third order quantities and adopting a
different approximation for $V_{td}$ and $V_{ts}$ in equations
(41-43), namely:
\begin{eqnarray}\label{eq: ndili51}
V_{td}& \approx & \sqrt{\frac{m_d}{m_s}}\left[  e^{i\phi}
\sqrt{\frac{C_d}{m_b}}\sqrt{(1- \frac{C_u}{m_t})} +
\sqrt{\frac{C_u}{m_t}}\sqrt{(1-\frac{C_d}{m_b})}\right]
\nonumber\\
  &  &
\end{eqnarray}
and
\begin{eqnarray}\label{eq: ndili52}
V_{ts}& \approx &   e^{i\phi}\sqrt{\frac{C_d}{m_b}}\sqrt{(1 -
\frac{C_u}{m_t})} +  \sqrt{\frac{C_u}{m_t}}\sqrt{(1 -
\frac{C_d}{m_b})} \nonumber\\
   &   &
\end{eqnarray}
we obtain :
\begin{equation}
\frac{|V_{td}|}{|V_{ts}|} = \sqrt{\frac{m_d}{m_s}}
\end{equation}
found in the 6 texture zero model. The fact remains however that
these re-approximations (49)-(53) are not compatible with the
approximation leading to equation (46). \\

\subsection{Determining the values of $C_u/m_t, C_d/m_b$}
We now take advantage of the intrinsic features of the 4 texture
zero model, contained in the several relations above, to determine
directly the quantities $C_u/m_t , C_d/m_b$. We use either of the
equations:
\begin{equation}
|V_{cs}| = 0.97341 =  \sqrt{\frac{C_u}{m_t}}\sqrt{\frac{C_d}{m_b}}
\end{equation}
or
\begin{equation}
 |V_{tb}| = 0.999135 = \sqrt{\frac{C_u}{m_t}}\sqrt{\frac{C_d}{m_b}}.
\end{equation}
as our basis to determine $C_u/m_t$ and $C_d/m_b$. We make the
simple assumption that  the relativity values $C_u/m_t$ and
$C_d/m_b$ are the same in the $M_u$ sector as in the $M_d$ sector
of texture zero mass matrices. This means we take  $C_u/m_t  =
C_d/m_b$ whence using $|V_{cs}|$ equation (37), we obtain:
\begin{equation}
\sqrt{\frac{C_u}{m_t}} = \sqrt{\frac{C_d}{m_b}} = \sqrt{0.97341} =
0.9866154
\end{equation}.
Then we deduce that:
\begin{equation}
\sqrt{(1 - \frac{C_u}{m_t})} = \sqrt{(1 - \frac{C_d}{m_d})} =
\sqrt{0.02659} = 0.1631.
\end{equation}
As input data for measured $V_{CKM}$ elements we have taken values
quoted in a recent paper by Lenz [15], namely:

\begin{equation}\label{eq: ndili64A}
\left(
\begin{array}{ccc}
|V_{ud}| & |V_{us}| & |V_{ub}|\\
|V_{cd}|  & |V_{cs}| & |V_{cb}| \\
|V_{td}| & |V_{ts}| & |V_{tb}|
\end{array}
\right) =  \left(
\begin{array}{ccc}
0.97426 \pm 0.00030 & 0.22545 \pm 0.00095 & 0.00356 \pm 0.00020\\
0.22529 \pm 0.00077  & 0.97341 \pm 0.00021 & 0.04508_{-0.00528}^{+0.00075} \\
0.00861_{-0.00037}^{+0.00021} & 0.04068 \pm 0.00138 &
0.999135_{-0.000018}^{+0.000057}
\end{array}
\right)
\end{equation}

For quark masses, we shall use the following values, with the
observation that the quark masses do in general run because of
renormalization group effects  which we do not consider here.

\begin{eqnarray}\label{eq: ndili63}
m_u  &=& 2.57 MeV  \nonumber\\
m_d  &=& 5.85 MeV  \nonumber\\
m_s  &=& 111.0 MeV  \nonumber\\
m_c (m_c) &=& 1.25 GeV  \nonumber\\
m_b (m_b) &=&  5.99 GeV   \nonumber\\
m_t  &=& 173.0 GeV
\end{eqnarray}

\subsection{The phase angles  $\phi_2$ and $\phi_1$}

We can next plug these quantities into the equation
\begin{eqnarray}\label{eq: ndili55}
|V_{cb}| & \approx & \left|\sqrt{\frac{C_u}{m_t}}\sqrt{(1 -
\frac{C_d}{m_b})} +
e^{-i\phi_2}\sqrt{\frac{C_d}{m_b}}\sqrt{(1-\frac{C_u}{m_t})}\right|
\end{eqnarray}
or
\begin{eqnarray}\label{eq: ndili56}
|V_{ts}| & \approx  & \left| \sqrt{\frac{C_d}{m_b}}\sqrt{(1 -
\frac{C_u}{m_t})} + e^{-i\phi_2} \sqrt{\frac{C_u}{m_t}}\sqrt{(1 -
\frac{C_d}{m_b})} \right|
\end{eqnarray}
and determine phase angle $\phi_2$ therefrom.  We use the
$|V_{cb}|$ equation and insert its measured value of $|V_{cb}| =
0.04508$. We now find the choice of angle $\phi_2$ for which the
right hand side yields the same numerical value as the data. We
get $\phi_2 = 164^o$. If we use the $|V_{ts}|$ equation whose
measured value is $ |V_{ts}| = 0.04068$ we get $\phi_2 = 165.5^o$. \\

We can next determine the phase angle $\phi_1$ from the two
equations:
\begin{eqnarray}\label{eq: ndili57}
|V_{us}| & \approx & \left|\sqrt{\frac{m_d}{m_s}} +
e^{-i\phi_1}\sqrt{\frac{m_u}{m_c}}\sqrt{\frac{C_u}{m_t}}\sqrt{\frac{C_d}{m_b}}
\right|
\end{eqnarray}
or
\begin{eqnarray}\label{eq: ndili58}
|V_{cd}| & \approx & \left|\sqrt{\frac{m_u}{m_c}} +
e^{-i\phi_1}\sqrt{\frac{m_d}{m_s}}\sqrt{\frac{C_u}{m_t}}\sqrt{\frac{C_d}{m_b}}
\right|
\end{eqnarray}

We obtain the following values: $\phi_1 = 100.6^o$ from $|V_{us}|
= 0.22545$, or $\phi_1 = 100.7^o$ from $|V_{cd}| = 0.22529$. Phase
angle $\psi = \phi_1 + \phi_2$, while angle $\phi = \phi_2$ from
equation (5). \\

\subsection{Other parameters of $M_u$ and $M_d$.}
Finally we find other parameters of the texture zero mass matrices
$M_u$ and $M_d$, by using the set of equations (6):
\begin{equation}\label{eq: ndili59}
|A_u|  =  \sqrt{\frac{m_u m_c m_t}{C_u}} = \sqrt{\frac{m_u
m_c}{C_u/m_t}} = 0.05744
\end{equation}

\begin{equation}
|B_u|  = \sqrt{\frac{(C_u-m_u)(C_u+m_c)(m_t-C_u)}{C_u}} \approx
\sqrt{m_tC_u(1 + \frac{m_c}{C_u})}\sqrt{(1-\frac{C_u}{m_t})} =
27.942
\end{equation}

\begin{equation}
D_u  =  m_u-m_c+m_t - C_u = m_t[\frac{m_u}{m_t} -\frac{m_c}{m_t} +
(1 - \frac{C_u}{m_t})] \approx m_t(1 - \frac{C_u}{m_t}) = 4.60
\end{equation}

\begin{equation}
|A_d|  =  \sqrt{\frac{m_d m_s m_b}{C_d}} = \sqrt{\frac{m_d m_s
}{C_d/m_b}} = 0.02583
\end{equation}

\begin{equation}
 |B_d| = \sqrt{\frac{(C_d-m_d)(C_d+m_s)(m_b-C_d)}{C_d}} \approx
  \sqrt{m_bC_d(1 + \frac{m_s}{C_d})}\sqrt{(1-\frac{C_d}{m_b})} =
0.9730
\end{equation}

\begin{equation}
 D_d  =  m_d-m_s+m_b - C_d  \approx m_b(1 - \frac{C_d}{m_b}) =
 0.1592
\end{equation}

Then in summary we have for the  4 texture zero mass matrices
$M_u, M_d$ that their parameters have the values (in GeV):
\begin{eqnarray}\label{eq: ndili58F}
M_u : C_u & = & 168.39993 ; |B_u| = 27.942 ; D_u = 4.60 ; |A_u| =
0.05744 \nonumber\\
M_d : C_d & = &  5.831 ;  |B_d| = 0.9730 ; D_d = 0.1592 ; |A_d| =
0.02583
\end{eqnarray}

 We see that these parameters have a definite hierarchy : $ C_u
\gg |B_u| \gg D_u \gg |A_u|$. Also $C_d \gg |B_d| \gg D_d \gg
|A_d|$. Systematically too, the $M_u$ parameters are larger in
value than the corresponding $M_d$ parameters. But superposed on
these features, is a striking hierarchical relationship among the
(2,2), (2,3) and (3,3) parameters in each of $M_u$ and $M_d$. The
relationship is that for each of $M_u, M_d$, we have : $|B_q|^2 =
C_q D_q$ where $ q = u, d$. This can be verified by our actual
parameter values above: $|B_u|^2 = 27.942^2 = 780.755$ against
$C_u D_u = (168.39993)(4.60) = 774.63$ for $M_u$. Also $|B_d|^2 =
0.9730^2 = 0.9467$ against $C_d D_d = (5.831)(0.1592) = 0.9283$
for $M_d$. This relationship has been called geometric hierarchy
by Fritzsch and Xing [1] who first noticed it. Our model
exhibits it closely. \\

But then, we notice other striking hierarchical feature of our
model when we transform our above parameter values into ratio
quantities, thus:
\begin{eqnarray}\label{eq: ndili58A}
|B_u|/C_u & = & 0.1659; |B_d|/C_d = 0.1668 \nonumber\\
D_u/|B_u| & = & 0.1646; D_d/|B_d| = 0.1636 \nonumber\\
|A_u|/D_u  &= & 0.01248 ; |A_d|/D_d = 0.1622
\end{eqnarray}
We see the ratios appear to obey equal spacing rule that is
particularly \\ noticeable in the $M_d$ case. The ratios of
succeeding elements, including the (1,2) or (2,1) element, are the
same across the mass matrix. This prompts further examination of
the hierarchy issue which we now  undertake. \\

\section{The hierarchy question in the 4 texture zero Model}

It is clear from the analysis above that the 4 texture zero model
as presented is characterized by certain distinctive hierarchies
among the parameters of the mass matrices $M_u$ and $M_d$. A
logical deduction is  that whatever success the model has in
explaining experimental data on flavor mixing and flavor physics
generally, must have as a factor, this dominant hierarchical state
or phase of the model. Then a question arises whether the model
has one unique hierarchical state that is in agreement with
experiment, or a number of such experimentally allowed
hierarchical states or phases. What in the end would such
duplicity of allowed hierarchical states of $M_u, M_d$ mean from a
symmetry point of view?  We find we can provide below some
answers and insights into these questions. \\

Our approach to the problem is to hold as central, the observed
equal spacing rule equation (72) found for $M_u$ and $M_d$, and
use it to characterize the 4 texture zero model. We assume as
reasonably exhibited in equation (72) that the rule applies
equally in the $M_u$ sector as in the $M_d$ sector. We denote the
value of the equal spacing by a number k where $ k = 0.16$ for
equation (72) that we now regard as just one of several possible
allowed hierarchical states or modes of the 4 texture zero model.
We show that similar states of $(M_u, M_d)$ exist that are also
dictated by present experimental data, but that each such other
state has a different k value that distinguishes it. Thus k is to
be seen as playing the role of a quantum number that labels
different hierarchical states of $(M_u, M_d)$, where each k-state
describes experimental data in much the same way as other k
labelled quantum states of $M_u$ and $M_d$. \\

Before discussing k further, we notice that our specification of
the parameters of $M_u$ and $M_d$ in equations (65) - (72) needed
a prior input data or specification for the quantities $C_u/m_t$
and $C_d/m_b$. This choice and specification we made in equation
(55), and this alone enabled us to calculate and fix the
parameters $C, |B|, D, |A|$ of the model for both $M_u$ and $M_d$.
We can say then that our 4 texture zero model was characterized
not only by k, but also by another quantum number we denote by N
and define as:
\begin{equation}
N = \frac{C_u}{m_t} = \frac{C_d}{m_b}
\end{equation}
Given N and k, we can specify completely the parameters of $M_u$
and $M_d$, It was a specific choice of $N = N_1 = |V_{cs}| =
0.97341$ in equation (55) that defined our initial hierarchical
state of $(M_u, M_d)$, with its $k = k_1 = 0.16$ given in equation
(72).\\

We can assert, on the above basis,  that the 4 texture zero model
through its mass matrices $M_u$ and $M_d$,  exists in hierarchical
quantum states that are described by just two mutually commuting
quantum numbers $(N, k)$. The quantum number N specifies the mass
scale of the system while the quantum number k specifies the rate
of fall off of the mass scale with flavor, for given N. \\

Our assertion is buttressed further by the fact that equation (56)
which is another valid experimental data like equation (55),
already shows that another choice of N value is possible given by
$N = N_2 = 0.999135 = |V_{tb}|$.  We can calculate the k number
associated with this $N_2$  and show that $k = k_2$ so obtained is
different from $k_1$ value of equation (72), thus defining another
allowed $(N, k)$ quantum state of $(M_u,M_d)$. We give these
details. \\

\subsection{The $(N_2, k_2)$ quantum state of $M_u$ and $M_d$}
In place of equation (55) for $N = N_1$ we choose  $N = N_2$ from
equation (56) given by:
\begin{equation}
N = N_2 = |V_{tb}| = 0.999135 =
\sqrt{\frac{C_u}{m_t}}\sqrt{\frac{C_d}{m_b}} = \frac{C_u}{m_t} =
\frac{C_d}{m_b}
\end{equation}
Then working our way through equations (66) -(72) as before, we
obtain the following values for the $N_2$ state:

\begin{eqnarray}\label{eq: ndili63A}
M_u : C_u &=& 172.850 ; |B_u| = 5.104238 ; D_u = 0.149645 ; |A_u| = 0.05670  \nonumber\\
M_d : C_d &=& 5.9848 ; |B_d| = 0.177720 ; D_d = 0.00518134 ; |A_d|
= 0.025493375 \nonumber\\
  &  &
\end{eqnarray}

whence :
\begin{eqnarray}\label{eq: ndili58F}
|B_u|/C_u & = & 0.02952981; |B_d|/C_d = 0.029695216 \nonumber\\
D_u/|B_u| & = & 0.029327794; D_d/|B_d| = 0.02915449 \nonumber\\
|A_u|/D_u & = & 0.378919777 ; |A_d|/D_d = 4.9
\end{eqnarray}

which defines a sustained equal spacing rule with a $k_2 \approx
0.0294 $ , though the rule appears visibly broken by the low mass
$A_u, A_d$ sector as we saw partially  in equation (72) for
$(N_1, k_1)$ state. \\

The phase angles computed for $M_u$ and $M_d$ in the $(N_2, k_2)$
state are:
\begin{eqnarray}\label{eq: ndili58E}
\phi_1 & = &  100.88^o from  :  |V_{us}| \nonumber\\
\phi_1 & = &  100.84^o from  :  |V_{cd}| \nonumber\\
\phi_2 & = &  79.88^o  from  :  |V_{cb}| \nonumber\\
\phi_2 & = &  92.44^o  from  :  |V_{ts}|
\end{eqnarray}
These angles are comparable to the ones found earlier for the
$(N_1, k_1)$ state, suggesting that the phase angles are not
suitable parameters or numbers to use to characterize $M_u$
and $M_d$. \\

We can generate other $(N_i, k_i), i, j = 1, 2, 3, 4, 5, ....$
quantum states of the $M_u, M_d$ system by stretching the
quantities $|V_{cs}|$ and $|V_{tb}$ used earlier to define $N_1$
and $N_2$, by their experimental error bars, in a manner suggested
by Xing and Zhang [3] and also Verma et. al. [4]. For example in
place of $|V_{cs}| = 0.97341 = N_1$ used in equation (55) to
define $N_1,$ we can choose from equation (59): $|V_{cs}| =
0.97362 = N_3$ and calculate its corresponding $k_3$ value thus
generating a new and still experimentally allowed quantum state
$(N_3 , k_3)$ of our $M_u, M_d$ system. Therefore our analysis
leads us to the conclusion that the flavor systems $M_u$ and $M_d$
live in a plurality of allowed quantum states $(N_i , k_i)$. \\

Independent investigations carried out by Yu-Feng Zhou[12] and by
Giraldo[17] indicate also that other variants of the 4 texture
zero model, not necessarily of equation (1) pattern, exhibit also
hierarchical features amenable to the $(N, k)$ quantum state
description or labelling for $(M_u, M_d)$.

\section{Possible symmetry import of the equal spacing rule}

It remains for us to attach some meaning, possibly of symmetry
nature,  to our finding of the above plurality of states $(N_i,
k_i)$ for $M_u$ and $M_d$, each state labelled by only two
mutually commuting  quantum numbers, and each state representing
some observed or observable state of the multi-flavor quark mass
system $M_u$ of $(u,c,t)$ system,  and $M_d$ of $(d, s,b)$.  We
argue that one way to immediately understand the result is to
recall the intimate connection between certain texture zero mass
matrices and weak basis transformations(WBT). It is known through
several authors [16-18] that hermitian quark mass matrices $(M_u,
M_d)$ possess in general the property known as Weak Basis (WB)
transformation property whereby given any one pair $(M_u, M_d)$,
one can generate several other pairs by merely applying some
unitary operator U thus: \\

$(M_u, M_d) \rightarrow U^\dagger (M_u, M_d) U = (M_u' , M_d')$\\
or
\begin{eqnarray}\label{eq: ndili58F}
 M_u & \rightarrow & M_u' = U^\dagger M_u U  \nonumber\\
M_d & \rightarrow & M_d' = U^\dagger M_d U
\end{eqnarray}

and that these different pairs $(M_u, M_d), (M_u', M_d'), (M_u",
M_d")......$ are equivalent with respect to the $V_{CKM}$ flavor
physics content of the system.   These matrices U constitute a
group we may call the WB group of transformations, and denote by
$G_{WB}$. In a general situation of arbitrary hermitian quark mass
matrices  $G_{WB}$  is the U(3) symmetry group. But in a situation
where we require or restrict all pairs $(M_u, M_d)$, $(M_u',
M_d')$, $(M_u" , M_d")$ ... to have the same 4 texture zero
patterns stated in our equations (1) and (2), we expect the
relevant WB group to be some subgroup of U(3).  The most
appropriate subgroup choice is SU(3), which we now show ties up
with our $(N_i, k_i)$ observed quantum states of $M_u, M_d$. \\

The fact is that the WB transformations of equation (78) imply
that each paired matrix $(M_u,M_d)$ is a representation  of
$G_{WB}$. Any such representation or pair, must carry some
commuting quantum numbers of $G_{WB}$ that label the
representation and distinguish one representation or pair from
another. Since the various pairs we actually computed from data
have each two commuting quantum number labels, and since SU(3) is
known to have two such commuting number labelling for its
representations, we conclude that our finding  can be explained as
the operation of WB symmetry in which SU(3) is specifically the WB
transformation group. What this throws up immediately is that in
the ongoing efforts[10,11] to embed the 4 texture zero model in a
GUT scheme and its low energy effective theory, we have to reckon
that our 4 texture zero model is subject to a plurality of $(N_i,
k_i)$ quantum number labelling. We will pursue this line of
thought elsewhere. \\

\section{Summary and Conclusions}
In conclusion, we have shown that the 4 texture zero quark mass
matrix model possesses strong hierarchical features which manifest
in a full blown equal spacing rule, or minimally a  geometric
hierarchy.  The features lead us to an important observation that
the quark mass matrices $(M_u, M_d)$ exist in a series of quantum
states  each labelled by two commuting numbers (N, k).  The two
numbers select SU(3) as a Weak Basis (WB) symmetry group of the
pair of mass matrices $(M_u, M_d)$. The two quantum numbers (N,k)
are expected to be relevant in ongoing efforts to embed the 4
texture zero model in SO(10) and other GUT theories of Particle
physics. \\

\textbf{References :} \\

[1]. H. Fritzsch and  Z. Z. Xing: Phys. Lett B555, 63 (2003). \\

[2]. H. Fritzsch and Z.Z. Xing: Prog. Part. Nucl. Phys. 45, 1 (2000) \\

[3]. Z.Z.Xing and H.Zhang: J. Phys. G30, 129 (2004) \\

[4]. R. Verma et. al: J. Phys. G. Nucl. Part. Phys.37, 075020 (2010). \\

[5]. M. Gupta, G. Ahuja, R. Verma, Int. J. Mod. Phys. A24S1, 3462
(2009) \\

[6]. P. S. Gill and M. Gupta, Phy. Rev. D56, 3143 (1997) \\

[7]. R. Verma, G. Ahuja, M. Gupta: Phys. Lett. B681, 330 (2009).
\\

[8]. G. Ahuja, M. Gupta, S. Kumar, M. Randhawa, Physw. Lett. B647,
394 (2007) \\

[9] M. Randhawa, V. Bhatnager, P.S. Gill, M. Gupta, Phys. Rev.
D60, 051301, (1999); arXiv.hep-ph/9903.428 \\

[10]. T. Fukuyama, Koich Matsuda, H. Nishiura, Int. J. Mod. Phys.
A22, 5325 (2007) \\

[11]. K. Matsuda,  T. Fukuyama, H. Nishiura, Phys. Rev. D61,
053001 (2000); hep-ph/9906433. \\

 [12]. Yu Feng Zhou, Textures and hierarchies in quark mass
matrices with four texture zeros, arXiv.hep-ph/0309076 (2003) \\

[13]. H. Fritzsch, Phys. Lett. B73, 317 (1978). \\

[14]. H. Fritzsch,  Nucl. Phys. B155, 189 (1979) \\

[15]. A. Lenz, Theoretical status of the CKM matrix,
arXiv.hep-ph/1108.1218 (2011) \\

[16]. G. C. Branco, D. Emmanuel Costa and R. Gonzalez Felipe.
Phys. Lett. B477, 147 (2000). \\

[17]. Y. Giraldo, Texture Zeros and WB Transformations in Quark
sector of the Standard Model, arXiv.hep-ph/1110.5986 (2011) \\

[18]. W. Ponce and R. Benavides. Euro. Phys. J. C71, 1641 (2011).
\\

\end{document}